\documentclass[sn-aps]{sn-jnl}

\jyear{2021}%

\usepackage{graphicx}
\begin{document}

\title[Article Title]{On the gravitization of quantum mechanics and wave function reduction in Bohmian quantum mechanics}


\author*[1,2]{\fnm{Faramarz} \sur{Rahmani}}\email{farmarz.rahmani@abru.ac.ir;faramarzrahmani@ipm.ir}
\affil*{Department of Physics, School of Sciences, Ayatollah Boroujerdi University, Boroujerd, Iran}
\affil[2]{School of Physics, Institute for Research in Fundamental Science(IPM), Tehran, Iran}
\author[2,3]{\fnm{Mehdi} \sur{Golshani}}\email{golshani@sharif.edu}
\affil[3]{Department of Physics, Sharif University of Technology, Tehran, Iran}
\abstract{The main topic of this paper is using Einstein's equivalence principle in the description of the gravity-induced wave function reduction in the framework of Bohmian causal quantum theory. However, such concept has been introduced and explored by Penrose for the standard quantum mechanics, but the capabilities of Bohmian quantum mechanics makes it possible to get some of results more clearly. In this regard, the critical mass for  transition from the quantum world to the classical world, the reduction time of the wave function and the temperature that corresponds to the Unruh temperature will be obtained by applying  Einstein's equivalence principle for the quantum motion of particle.}




\maketitle
\section{Introduction}\label{sec:1}
One of the mysterious problems of quantum mechanics, which is not fully understood, is the problem of wave function reduction. We cannot predict the final state of the system after the reduction or collapse of the wave function in a deterministic manner. Rather, we are faced with a mixture of outcomes that we can only talk about the probability of the occurrence of a specific state. A fundamental question is how this processes occurs? Also we can ask, what is the objective criterion for determining the boundary between the quantum and classical worlds? In other words, how can we determine whether an object is a macroscopic or microscopic object?. As we know, the mass of a particle can be considered as a criterion to determine the boundary between the quantum and classical behaviors of a particle. The macroscopic objects obey the Newton laws of motion, while the dynamics of a microscopic object is governed by the Schr\"{o}dinger equation. In this regard, we need a gravitational theory which involves the mass of the particle as a criterion for transition from the quantum to classical world. This is done in the framework of gravity-induced wave function reduction in which the self-gravity of the particle or the spacetime curvature due to the particle mass, reduces the wave function of the particle \cite{RefK1,RefK2,RefK3}. Here, the self-gravity of the particle is justified through the concept of probability distribution in quantum mechanics. We know that a particle is detected around the point $\mathbf{x}$ in its configuration space with the probability density $\rho (\mathbf{x},t)=\vert \psi \vert ^2$ at time $t$. 
\begin{figure}[h] 
\centerline{\includegraphics[width=6cm]{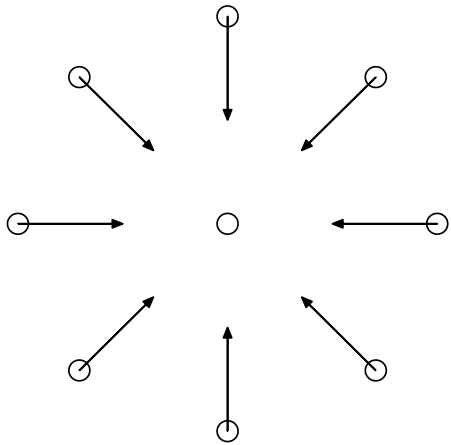}}
\caption{A schematic plot for the concept of self-gravity in configuration space. The probability distribution of particle is $\rho=\vert \psi \vert^2$ in this space. \label{fig:1}}
\end{figure}
Thus, the distribution of a point particle in the configuration space seems like an extended mass distribution. Hence, the definition of self-gravity is possible for a point particle in a quantum mechanical sense. In the following, we point to two of the most important approaches in the context of gravity-induced wave function reduction.\par 
One of the main approaches is based on the Schr\"{o}dinger-Newton equation \cite{RefD1,RefD2,RefD3}. A Gaussian wave packet with the initial width $\sigma_0$, spreads out over time(quantum mechanical behavior). To have a stationary wave packet, the particle mass  must be equal to a specific critical value in order to provide the required self-gravity to inhibit the dispersion of the wave packet.
The Schr\"{o}dinger-Newton equation for a single particle with distribution $\rho = \vert \psi(\mathbf{x},t) \vert ^2$ is:
\begin{equation}\label{sn}
i\hbar \frac{\partial\psi(\mathbf{x},t)}{\partial t}=\left(-\frac{\hbar^2}{2m}\nabla^2 -Gm^2 \int \frac{\vert \psi(\mathbf{x}^\prime,t)\vert^2}{\vert \mathbf{x}^\prime -\mathbf{x} \vert} d^3 x^\prime\right) \psi(\mathbf{x},t).
\end{equation}
Minimizing the Hamiltonian functional equation of the Schr\"{o}dinger-Newton equation for a stationary wave packet, leads to a relation between the critical mass of the particle and the characteristic width of its associated stationary wave packet \cite{RefD1}. In other words, the value of $\sigma_0$ for which the  wave packet remains stationary, is determined as follows:
\begin{equation}\label{Dio}
\sigma_{0} = \frac{\hbar^2}{Gm^3}.
\end{equation}
The width of the wave packet on the left hand side of this equation is related to the objective quantities on the right hand side. This enables us to determine the characteristic width of the wave packet objectively. By using  the relation (\ref{Dio}), a critical mass for transition from the quantum world to classical world is defined. It is given by
\begin{equation}\label{mc}
m_c=\left(\frac{\hbar^2}{G\sigma_0 }\right)^{\frac{1}{3}}.
\end{equation}
With a fixed $\sigma_0$, the particles with masses greater than the critical mass represent more macroscopic behavior, and for the particles with masses less than the critical mass, micro behaviors increase. The different classifications of particle motion in reduction processes in Bohmian context has been investigated in Ref \cite{RefRGG3}. Also, a basic Bohmian explanation for gravity-induced wave function reduction can be found in Refs \cite{RefRGG1,RefRGG2}. In practice, we must consider an object with a definite size. For an object with the ordinary matter densities, the critical mass is of the order of the Planck mass $m_p \approx 10^{-8}Kg$. See Refs \cite{RefK1,RefBassi}. To study the gravitational reduction of the wave function in the framework of the standard quantum mechanics, Refs \cite{RefGu,Refm1,Refde} are suggested. \par 
Using Einstein's equivalence principle in the study of wave function reduction was introduced by Penrose  under the title of "gravitization of quantum mechanics" \cite{RefP1,RefP2,RefP3}. 
It means bringing quantum mechanics closer to the principles of general relativity. We know that the theory of gravity is a geometric, intuitive, and visualizable local theory. But in quantum mechanics, we encounter non-classical and non-local behaviors such as entanglement, superposition of states, etc. Specially, in the standard quantum mechanics, the concept of trajectory  and pre-measurement quantities disappear and the physical quantities are defined as operators in a Hilbert space. On the other hand, one of the approaches in which an attempt is made to have a realistic description of quantum mechanics, is the causal quantum theory of de Broglie-Bohm. \cite{Refdeb,RefB,RefI,RefU,RefH,RefCush}. But, as Bohm himself said, this is not the last word \cite{RefCB}. Rather, it shows that a causal interpretation of quantum mechanics is not impossible. Some of the concepts that have no room in the standard quantum mechanics, can be defined in a more clear way in the Bohmian quantum mechanics. An example, is the concept of particle motion in quantum domain. This, motivated us to study the wave function reduction by applying the principle of equivalence for the quantum motion of particle in order to provide a clear picture of the problem.\par 
First, we take a look at the concept of "gravitization of quantum mechanics" for the problem of wave function reduction. Due to the universality of quantum mechanics, we expect that an object to be in a superposition of different states of position simultaneously. But, we do not observe such a thing for macroscopic objects. In the gravity-induced approaches, the reason of breaking the superposition of different  states of the particle is the self-gravity of the particle.\par
In Ref \cite{RefP3}, it has been argued that if we use the principles of general relativity, the superposition of the quantum states of the particle at different locations, in the presence of gravity, is not stable. Because it leads to different vacua and the superposition of different spacetimes is illegal. To clarify the effect of gravity, consider an object in two different locations with their associated states $\vert \phi_i \rangle, i=1,2$. Each state satisfies the Schr\"{o}dinger equation separately as a stationary state, with a unique Killing vector. The superposed state 
\begin{equation}
\vert \psi \rangle = \alpha \vert \phi_1\rangle +  \beta \vert \phi_2 \rangle 
\end{equation}
 is also a stationary state with the unique Killing vector $\mathcal{K}=\frac{\partial}{\partial t}$. When, the self-gravity of the object (the spacetime curvature due to the mass of the object) is taken into account, the total quantum state becomes
 \begin{equation}\label{ghh}
 \vert\psi \rangle_{\mathcal{G}}= \alpha \vert \phi_1\rangle \vert \mathcal{G}_1 \rangle +  \beta \vert \phi_2\rangle \vert \mathcal{G}_2 \rangle
\end{equation}  
where, $\vert \mathcal{G}_i \rangle, i=1,2$, are the quantum state of the gravity in the location of the particle. But, $\vert\psi \rangle_{\mathcal{G}}$ is not a stationary state in the sense that it has a unique Killing vector. Consequently, it decays to a stable state. The lifetime of the superposition is
\begin{equation}\label{lt}
\tau \approx \frac{\hbar}{\Delta E_G}
\end{equation}
where, $\Delta E_G$ denotes the uncertainty in the gravitational self-energy of the mass distributions of the two stationary states. This interesting result is a consequence of the concept of gravitization of quantum mechanics. For details see\cite{RefP3}. Practically, we cannot predict the final state of the system exactly. Because, we do not have enough information to make this prediction. In other words, the unitary evolution of the Schr\"{o}dinger equation breaks down during the reduction of the wave function. Finally, we only have a mixture of states. \par
The measurement problem also is justified through the gravity-induced wave function reduction. In fact, the self-gravity of the measuring apparatus as a macroscopic object, reduces the wave function of the measuring apparatus. Because, its wave function  is entangled with the different states of the quantum system. Consequently, the wave function of the system following the reduction of the wave function of the apparatus, reduces to a specific state. Schr\"{o}dinger's cat is also justified in the same way. \cite{RefP3}. \par 
In the Bohmian quantum mechanics there is room for studying the motion of a particle in a causal manner. Bohmian quantum mechanics assumes that the particle motion is deterministic in principle, but we can only have statistical predictions and outcomes in measurements. In the Bohmian quantum mechanics we can talk about the probability that at the time $t$  particle \textit{lies} in the volume element $d^3\mathbf{x}$ around the point $\mathbf{x}$. While, in usual quantum mechanics we can only discuss about the \textit{detecting} of particle at the time $t$ in the volume element $d^3\mathbf{x}$ around the point $\mathbf{x}$. In other words, the deterministic evolution of a system is  basically possible in this theory. The quantum motion of a particle can be described by the modified Hamilton-Jacobi equation 
\begin{equation}\label{hamilton}
\frac{\partial S(\mathbf{x},t)}{\partial t}+\frac{(\nabla S)^2}{2m}+V(\mathbf{x})+Q(\mathbf{x})=0.
\end{equation}
Here, quantity $Q$ denotes the quantum potential of the particle and it is given by
\begin{equation}\label{qq}
Q=-\frac{\hbar^2 \nabla^2 R}{2mR}=-\frac{\hbar^2 \nabla^2 \sqrt{\rho}}{2m \sqrt{\rho}}.
\end{equation}
Where, $ \rho=\psi^{*}\psi =R^2$ is the probability distribution of the particle. Equation (\ref{hamilton}) can be obtained through the substitution of the polar form of the wave function, $\psi=Re^{\frac{iS}{\hbar}}$, into the Schr\"{o}dinger equation.
The energy of the system is related to the principal function or the action of the particle as
\begin{equation}\label{e1}
E=-\frac{\partial S}{\partial t}
\end{equation}
The position of the particle is described by the following equation:
\begin{equation}\label{guidance}
\frac{d\mathbf{x}(t)}{dt}=\left(\frac{\nabla S(\mathbf{x},t)}{m}\right)_{\mathbf{X}=\mathbf{x}(t)}=\left(\frac{\mathbf{p}}{m}\right)_{\mathbf{x}=\mathbf{x}(t)}
\end{equation}
where $\mathbf{p}=\nabla S(x,t)$ is the momentum of the particle. The meaning of $\mathbf{x}=\mathbf{x}(t)$, is that the particle moves along a trajectory of the ensemble which is described by $\mathbf{x}(t)$.
Newton's second law for the quantum motion of the particle in an external potential $U$, takes the form
\begin{equation}\label{dn}
\frac{d}{dt}(m\dot{\mathbf{x}})=-\nabla U-\nabla Q \vert_{\mathbf{x}=\mathbf{x}(t)}
\end{equation}
Here, $-\nabla Q$, represents the quantum force. In this theory, the quantum force and the quantum potential, give non-classical features to the system. Thus, we cannot say Bohmian quantum mechanics is a return to the classical world. According to the abilities of Bohmian quantum mechanics in describing the quantum motion of a particle, we expect that the application of Einstein's equivalence principle for the problem of wave function reduction in Bohmian context, gives clear and interesting results. However, the use of equivalence principle for this problem has been done in usual quantum mechanics too, but we shall see that there are some points and results that are not easy to achieve in the framework of the standard quantum mechanics directly. In the following, we first discuss about the WEP in the Bohmian quantum mechanics.
\section{Einstein's equivalence principle in Bohmian quantum mechanics}\label{sec:2}
In classical mechanics, the weak equivalence principle of general relativity(WEP) is expressed in different  ways, all of which are basically the same. In the following, we shall argue that all statements are not equivalent in the Bohmian quantum mechanics. Three of the most famous statements of WEP in classical physics are as follows.\\ 
\textbf{The first statement}: \textit{Inertial mass is equivalent to passive gravitational mass: $m_i=m_g$}.\\
\textbf{The second statement}: \textit{The behavior of a freely-falling test particle is universal: $\mathbf{a}=-\mathbf{g}$}.\\
\textbf{The third statement}: \textit{In small enough regions of spacetime, the motion of freely-falling bodies are the same in a gravitational field and in a uniformly accelerated frame}.\par 
Before we start the discussion, let us make a point about the inertial and gravitational masses. The inertial mass $m_i$, in the first statement, has a universal character. Because it is defined in terms of resistance to momentum change by other forces and it does not matter what kind of force is exerted to it. On the other hand, $m_g$ is a quantity specific to the gravitational force. It can be thought of as a gravitational property or "gravitational charge".
In this work, we assume that the gravitational property $m_g$ is equal to the universal quantity $m_i$.  In other words, the extension of the first statement to quantum mechanics is unobjectionable.\par 
Now let's look at the second statement of WEP in the Bohmian quantum mechanics. The quantum version of Newton's second law in Bohmian quantum mechanics in the gravitational potential $U=m\mathbf{g}\cdot \mathbf{x}$ is:
\begin{equation}\label{d1}
\frac{d}{dt}(m\dot{\mathbf{x}})=-m\mathbf{g}-\nabla Q \vert_{\mathbf{x}=\mathbf{x}(t)}
\end{equation}
which leads to the equation
\begin{equation}\label{d2}
\ddot{\mathbf{x}}=-\mathbf{g}-\frac{1}{m}\nabla Q \vert_{\mathbf{x}=\mathbf{x}(t)}
\end{equation}
Equation (\ref{d2}) shows the violation of the second statement explicitly; even with the equality of gravitational and inertial masses. Therefore, the first and second statements are not equivalent in Bohmian quantum mechanics. In general, the quantum potential $Q$ depends on the mass of the particle. Thus, the violation of the second statement in (\ref{d2}) is not only due to the $\frac{1}{m}$ coefficient. We shall use relation (\ref{d2}) to study the particle motion under the influence of  its own gravity and quantum force. \par 
Checking the validity of third statement has been widely done by some authors \cite{RefP1,RefNa}. Consider a non-relativistic falling object in a homogeneous gravitational field $\mathbf{g}$ with the coordinate system $(\mathbf{x},t)$. The Schr\"{o}dinger equation for this system is
\begin{equation}\label{ns}
i\hbar \frac{\partial\psi(\mathbf{x},t)}{\partial t}=-\frac{\hbar^2}{2m}\nabla^2 \psi -m\mathbf{x}\cdot \mathbf{g}\psi(\mathbf{x},t).
\end{equation}
According to the third statement, we can change the coordinate systems by using the transformations
\begin{equation}\label{trans}
\begin{aligned}
\mathbf{x}^{\prime}&=\mathbf{x}-\frac{1}{2}\mathbf{g}t^2 \\
t^{\prime}&=t
\end{aligned}
\end{equation}
to get the Schr\"{o}dinger equation in the accelerated frame with the acceleration  $\mathbf{g}$. According to the third statement, an observer in this accelerated frame, feels no sense of gravity. For this observer, the Schr\"{o}dinger equation takes the form of the free Schr\"{o}dinger equation which is given by
\begin{equation}\label{rs}
i\hbar \frac{\partial\phi}{\partial t^{\prime}}=-\frac{\hbar^2}{2m}\nabla^2 \phi.
\end{equation}
Here, $\phi$ is the wave function of the freely-falling particle in this frame. The functions $\phi$ and $\psi$ are called "Einsteinian" and "Newtonian" wave functions\cite{RefP3}. To establish the  third statement of WEP, two wave functions $\psi$ and $\phi$ must be related in the form
\begin{equation}\label{wr}
\phi(\mathbf{x}^{\prime},t^{\prime})=\exp\left(\frac{im}{\hbar}\left(\frac{\mathbf{g}^2 t^3}{6}-\mathbf{x}\cdot \mathbf{g} t\right)\right)\psi(\mathbf{x},t)
\end{equation}
or equivalently ,
\begin{equation}\label{wr2}
\psi(\mathbf{x},t)=\exp\left(\frac{im}{\hbar}\left(\frac{\mathbf{g}^2 t^{\prime ^3}}{3}+\mathbf{x}^{\prime}\cdot \mathbf{g} t^{\prime}\right)\right)\phi(\mathbf{x}^{\prime},t^{\prime}).
\end{equation}
These relations are also valid in the Bohmian quantum mechanics with the difference that it is possible to use the phase of theses relations to study the energy and dynamics of the particle. While, this is not possible in the standard quantum mechanics. We shall see this in the next section.
However, these relations are derived for a particle in a homogeneous gravitational field and are the consequence of the establishment of equivalence principle in quantum mechanics, but we can use these relations in the problem of  wave function reduction  where $\mathbf{g}$ denotes  the self-gravitational field of the particle and can be considered approximately homogeneous in a short-time estimation.\par 
In Ref \cite{RefP3} it is argued that the nonlinear term $\frac{m \mathbf{g}^2 t^3}{\hbar} $ breaks down the notion of positive frequency and this refers to the different vacua or spacetimes. Since, the superposition of different vacua is illegal, such superposition must be reduced to a stationary state during the decay time $\tau \approx \frac{\hbar}{\Delta E_G}$ through a non-unitary evolution. Finally, we are faced with a mixture of states. This is what also happens in quantum field theory in curved spacetime or quantum field theory in an accelerated frame. There, a pure vacuum state of a quantum field for an inertial observer, is seen as a mixture of states for an accelerated observer. The accelerated observer detects particles in his/her vacuum with the Unruh temperature $T=\frac{\hbar a}{2\pi c k_B}$ where $a$ denotes the acceleration of the observer or frame.\cite{RefPad,RefCm,RefUW}. This shows that there is the same physical concept behind the both phenomena which manifests itself as a transformation from pure quantum state to the mixture of states. We shall show that in the context of the Bohmian quantum mechanics, a similar temperature can be attributed systematically to the particle distribution in configuration space. But, we first show how applying the principle of equivalence to the quantum motion of particle leads to the criterion (\ref{Dio}) in Bohmian context.
\section{The motion of particle and wave function reduction}  
Now, we want to investigate the problem of wave function reduction in  Bohmian quantum mechanics for a particle moving in its own gravity. First, we review the quantum motion of a particle in an external  homogeneous gravitational field. The  motion of a particle in its own gravity can be studied by the same equations in a short-time estimation. In a short-time estimation the self-gravitational field of the particle and the width of its associated wave packet are approximately  constant\cite{RefRGG3}.
 By using the contents of Ref \cite{RefH}, a particle falling in a constant gravitational field with zero initial velocity is guided by a Gaussian wave packet given by:
\begin{equation}\label{g1}
\begin{aligned}
\psi(\mathbf{x},t)&=& \\ &\left(2\pi \sigma_0^2\left(1+\frac{i\hbar t}{2m\sigma_0^2}\right)^2\right)^{-\frac{3}{4}} \exp\left\lbrace -\frac{(\mathbf{x}+\frac{1}{2}\mathbf{g}t^2)^2}{4\sigma_0^2\left(1+\frac{i\hbar t}{2m\sigma_0^2}\right)} +\frac{im}{\hbar}(\mathbf{x}^2-\mathbf{g}\cdot \mathbf{x}t -\frac{1}{6}m \mathbf{g}^2t^3)\right\rbrace
\end{aligned}
\end{equation}
The amplitude  of the wave packet is as follows. 
\begin{equation}\label{amp}
R=(2\pi \sigma^2)^{-\frac{3}{4}} \exp \left\lbrace -\frac{(\mathbf{x}+\frac{1}{2}\mathbf{g}t^2)^2}{4\sigma^2} \right\rbrace
\end{equation}
Here, $\sigma$ denotes the width of the wave packet at time $t$, and it is given by:
\begin{equation}
\sigma=\sigma_0 \sqrt{1+\frac{\hbar^2 t^2}{4m^2\sigma_0^2}}
\end{equation}  
The width $\sigma$ of the wave packet in a homogeneous gravitational field is the same as the width of a free wave packet\cite{RefH}. In a short-time estimation, $\sigma \approx \sigma_0$. Also, the probability distribution $\rho$ can be considered Gaussian with a good approximation \cite{RefRGG3}. \par 
It can be shown that the acceleration of the particle is given by:
\begin{equation}\label{ac2}
\ddot{\mathbf{x}}=-\mathbf{g}+\frac{\hbar^2 \mathbf{x}_0}{4m^2\sigma_0\sigma^3}
\end{equation}
Where, the first term is gravitational acceleration and the second term denotes the quantum or Bohmian acceleration.\par 
Now, we discuss about the reduction of the wave function by using the third statement of WEP. 
In general, the acceleration of a particle in an accelerated frame with the acceleration $\mathbf{a}$ and  coordinates $(\mathbf{x}^\prime,t^\prime)$, in a homogeneous gravitational field is
\begin{equation}
\ddot{\mathbf{x}}^\prime =\mathbf{a}-\mathbf{g}
\end{equation}
In order to establish the third statement of (WEP), we must have $\ddot{\mathbf{x}}^\prime=0$. In other words, the acceleration of the frame must be equal to the gravitational acceleration to have a free motion in this frame.
On the other hand, there is an interesting and subtle point about the nature and the feature of Bohmian quantum force and that is the fact that quantum force is a type of inertial forces. In other words, when  the quantum force on a particle is considered, the dynamics of the particle is equivalent to the dynamics of a particle in a non-inertial frame with the quantum acceleration \cite{RefShoja,RefMach,RefVig}. This means that relation (\ref{ac2}) describes the acceleration of a particle in a non-inertial frame with the Bohmian or quantum acceleration. Of course quantum acceleration is not constant in general. But, in our discussion we need the average of the accelerations to obtain a criterion for the reduction of the wave function and the average of  quantum and self-gravitational accelerations in a short-time estimation are constant with a good approximation \cite{RefRGG3}.\par 
Here are some interesting points. Quantum force only vanishes for a plane wave. In a finite region of spacetime, specially in quantum world, we cannot have a plane wave. In other words, in relation (\ref{ac2}), we are always faced with a quantum acceleration. Thus, to have a free motion in a finite region of spacetime in an accelerated frame with the quantum acceleration, we must equalize gravitational and quantum accelerations. In fact, the quantum force is different from usual forces in physics, such as the electric force between two charges, the friction force and etc. Furthermore, the inertial forces do not obey Newton's third law\cite{RefIn}. Interestingly, this is one of the features of  Bohmian quantum force. In other words, it is a one-way force that is not related to any energy field. The derivation of the Bohmian quantum force as a non-inertial force under special conditions has been discussed in Ref \cite{RefMach}.\par 
 Now, to have a free motion in this accelerated frame (establishing of third statement of WEP), quantum acceleration and gravitational acceleration must be equal. But, to get an objective criterion we must obtain the averages of these accelerations or forces. The quantum acceleration and gravitational acceleration are the derivatives of the quantum potential and self-gravitational potential which are defined as follows \cite{RefRGG3}.
\begin{equation}\label{sq}
Q=-\frac{\hbar^2}{2m}\frac{\nabla^2 R}{R}= -\frac{\hbar^2}{2m}\frac{1}{r}\frac{\partial}{\partial r}\left(r\frac{\partial R}{\partial r}\right)=\frac{\hbar^2}{8m\sigma_0^4}\left(6\sigma_0^2-r^2\right)
\end{equation}
\begin{equation}\label{pot2}
U=- \int_{0}^{r} \frac{Gm^2}{r^{\prime}}\rho(r^{\prime}) dv^{\prime}=\sqrt{\frac{2}{\pi}}\frac{Gm^2}{\sigma_0}\left(1-e^{-\frac{r^2}{2\sigma_0^2}}\right).
\end{equation}
The volume element is $dv^{\prime}=4\pi r^{\prime^2} dr^{\prime}$. Here, for simplicity, we have written relations (\ref{sq}) and (\ref{pot2}) in spherical coordinates. Relation (\ref{pot2}), in a general form, is
\begin{equation}
U(\mathbf{x},t)=-Gm^2 \int \frac{\vert \psi(\mathbf{x}^\prime,t)\vert^2}{\vert \mathbf{x}^\prime -\mathbf{x} \vert} d^3 x^\prime
\end{equation}
which is the second term on the right hand side of relation (\ref{sn}). Relation (\ref{pot2}) shows its spherical form. In spherical coordinates, the amplitude of the wave packet (\ref{amp}), in a short-time estimation, takes the form
\begin{equation}
R(r)= (2\pi \sigma_0^2)^{-\frac{3}{4}}e^{-\frac{r^2}{4\sigma_0^2}}
\end{equation}
For details, see Ref \cite{RefRGG3}.
To obtain the average of quantum and gravitational accelerations, we can use relations
\begin{equation}\label{abar}
\vert\bar{\mathbf{a}}_q\vert=\frac{\vert\bar{\mathbf{f}}_q\vert}{m}=\frac{1}{m} \left(\vert - \int_0^{\infty} \rho(r) \nabla Q dv \vert \right)=\sqrt{\frac{2}{\pi}}\frac{\hbar^2 }{2m^2 \sigma_0^3}\approx \frac{\hbar^2 }{m^2 \sigma_0^3}
\end{equation} 
and 
\begin{equation}\label{gbar}
\bar{\vert \mathbf{g}\vert}=\frac{\vert\bar{\mathbf{f}_g}\vert}{m}=\frac{1}{m}\left(\vert - \int_0^{\infty} \rho(r) \nabla U  dv\vert \right) = \frac{1}{\pi}\frac{Gm}{\sigma_0^2} \approx \frac{Gm}{\sigma_0^2}.
\end{equation}
Where, $\rho=\psi^{*} \psi=R^2$. Here, $\mathbf{f}_q=-\nabla Q$ and $\mathbf{f}_g=-\nabla U$ are the quantum force and self-gravitational force respectively. For this distribution, the quantum force is
\begin{equation}
\mathbf{f}_q=-\nabla Q=\frac{\hbar^2 r}{4m\sigma_0^2}
\end{equation}
and the self-gravitational force takes the form
\begin{equation}
\mathbf{f}_g=-\nabla U=-\frac{\sqrt{2}\, \mathit{Gm}^{2} r \,{\mathrm e}^{-\frac{r^{2}}{2 \sigma_0^{2}}}}{\sqrt{\pi}\, \sigma_0^{3}}
\end{equation}
 By equating the average values of these forces or accelerations, we get
\begin{equation}\label{dio2}
 \sigma_0 = \frac{\hbar^2}{Gm^3}.
\end{equation}
This is the same as the famous criterion (\ref{Dio}) of Diosi . But, here, this relation is obtained through the investigation of WEP in Bohmian mechanics.  \par 
In the Ref \cite{RefRGG3}, by investigating the deviation equation for an ensemble of Bohmian trajectories, it has been shown  that there are three regimes for the particle motion under the influence of its own gravity and quantum force. For masses greater than the critical mass i.e. the gravity-dominant regime, gravitational force overcomes the quantum force and the motion of a particle with the Gaussian distribution in its own gravity in the spherical coordinates is described by equation
\begin{equation}\label{eqg}
r(t)=r(0) -\frac{1}{2}\frac{Gm}{\sigma_0^2}t^2
\end{equation}
For masses less than the critical mass, i.e. the quantum dominant regime, the quantum force overcomes the self-gravitational force and quantum mechanical behaviors increase. For masses equal to the critical mass, the stationary states appear, and this is called transition regime.
Relation (\ref{eqg}) can be used to obtain the time it takes for the particle to fall from $r(0)=\sigma_0$ to $r(\tau)=0$. It is given by
\begin{equation}\label{tt}
\tau= \left(\frac{\sigma_0^3}{Gm}\right)^{\frac{1}{2}}.
\end{equation}
Also, it has been shown in Ref \cite{RefRGG3} that relation (\ref{tt}) is equal to the decay time in relation (\ref{lt}) as was proposed by Penrose. Now, by using the criterion (\ref{dio2}), the relation (\ref{tt}) can be written completely objectively as follows.
\begin{equation}\label{tt2}
\tau= \frac{\hbar^3}{G^2m^5}.
\end{equation}
As we expect, the reduction time is proportional to the inverse of particle mass.
In the next section, we shall get relation (\ref{tt2}) through  investigating the difference in particle energy for two free and accelerated observers, and using the uncertainty relation. We shall also derive a relation for temperature that is mentioned in Ref \cite{RefP3}. 
 \section{Einsteinian and Newtonian observers in Bohmian quantum mechanics}
In Ref \cite{RefP3},  the decay time of a superposed stat consisting of two position states in the presence of gravity  is related to the uncertainty in the self-gravitational energy of the particle. Now, we want to relate this uncertainty to the difference in particle energy for  two Einsteinian and Newtonian observers(frames). Then, by using the uncertainty principle we shall estimate the reduction time. In Ref \cite{RefP3}, it has been pointed out that the accelerated observer should observe a thermal vacuum with a temperature corresponding to known Unruh temperature. But, in the non-relativistic regime ($c \rightarrow \infty$), the temperature goes to zero. In this section we show that this temperature is indeed proportional to the Unruh temperature. This is a new work in Bohmian context and has not been done before.\par
Let us to obtain the relation between the Einsteinian and Newtonian  wave functions in an accelerated frame with coordinates $(\mathbf{x}^{\prime} ,t^{\prime})$. For this purpose, we can use  relations (\ref{trans}), (\ref{wr}) and (\ref{wr2}) or refer to Ref  \cite{RefNa}. Any way, the desired relation is as follows.
\begin{equation}\label{w}
 \psi(\mathbf{x}^{\prime},t^{\prime})=\phi(\mathbf{x}^{\prime},t^{\prime}) e^{ \frac{i}{\hbar}(-m\mathbf{g}\cdot\mathbf{x}^{\prime}t^{\prime}+\frac{1}{3}m\mathbf{g}^2 t^{\prime 3})}
\end{equation} 
On the other hand, the Einsteinian wave function(free wave function) can be represented  as follows.
\begin{equation}
\phi(\mathbf{x}^{\prime},t^{\prime}) =\phi_0 e^{\frac{i}{\hbar}S(\mathbf{x}^{\prime},t^{\prime})}
\end{equation}
Now, equation (\ref{w}) takes the form
\begin{equation}\label{w2}
\psi(\mathbf{x}^{\prime},t^{\prime})=\phi_0 e^{\frac{iS(\mathbf{x}^{\prime},t^{\prime})}{\hbar}+\frac{i}{\hbar}(-m\mathbf{g}\cdot\mathbf{x}^{\prime}t^{\prime}+\frac{1}{3}m\mathbf{g}^2 t^{\prime 3}) } 
\end{equation}
where, $S(\mathbf{x}^{\prime},t^{\prime})$ is the free or the Einsteinian phase of the wave function. The Newtonian phase of the  wave function is
\begin{equation}\label{pha}
S^{\prime}(\mathbf{x}^{\prime},t^{\prime})=S(\mathbf{x}^{\prime},t^{\prime})-m\mathbf{g}\cdot \mathbf{x}^{\prime}t^{\prime}+\frac{1}{3}m\mathbf{g}^2 t^{\prime 3}
\end{equation}
The energy of the particle is
 \begin{equation}\label{en1}
 E^{\prime}=-\frac{\partial S^{\prime}(\mathbf{x}^{\prime},t^{\prime})}{\partial t^{\prime}}=-\frac{\partial S(\mathbf{x}^{\prime},t^{\prime})}{\partial t^{\prime}} +m\mathbf{g}\cdot \mathbf{x}^{\prime}-m\mathbf{g}^2t^{\prime 2}
 \end{equation}
Now, by using $\mathbf{x}^{\prime}=-\frac{1}{2}\mathbf{g}t^{\prime 2}$ and $t^{\prime}=t$, equation(\ref{en1}) becomes
\begin{equation}\label{en2}
E^{\prime}=-\frac{\partial S(\mathbf{x}^{\prime},t^{\prime})}{\partial t^{\prime}} -\frac{3}{2} m\mathbf{g}^2t^{2}=E-\frac{3}{2} m\mathbf{g}^2t^{ 2}
\end{equation}
The energy of the particle for the Newtonian observer is $E^{\prime}$, while for the Einsteinian one is $E$. Relation (\ref{en2}), shows that energy is not conserved and varies over time. In other words, the difference in particle energy for two observers is 
\begin{equation}\label{del}
\mathcal{E}(t) =E^{\prime}-E=-\frac{3}{2} m\mathbf{g}^2t^{ 2}
\end{equation}
It is clear that in the absence of gravitational field ($\mathbf{g}=0$), or in a flat spacetime, $\mathcal{E}=0$ and the Newtonian observer is the same as the Einsteinian one. This energy difference is due to the establishment of principle of equivalence. Because, this energy difference is obtained through the differentiation of the phase of the wave function with respect to time in relation (\ref{w2}). The form of this phase is a consequence of the establishment of Equivalence principle in quantum mechanics.\par
Now, to estimate the collapse time, we return to the wave packet and its Gaussian distribution in spherical coordinates. We assume that at $t=0$, the position of the particle is  $r(0)=\sigma_0$, and at $t=\tau$, the particle is at $r=0$. Then, during the reduction time $\tau$, a sphere with the radius $\sigma_0$ is formed in the configuration space. The uncertainty in  energy $\varepsilon(t)$ at $t=\tau$ must satisfy the relation 
\begin{equation}
\mathcal{E}\vert_{t=\tau} \tau \approx \hbar
\end{equation}
where by using relation (\ref{del}) we get 
\begin{equation}\label{1a}
m\mathbf{g}^2 \tau ^3 \approx \hbar
\end{equation}
Now, substituting relations (\ref{gbar}) and (\ref{dio2}) into relation(\ref{1a}), gives:
\begin{equation}
\tau=\frac{\hbar^3}{G^2	m^5}
\end{equation}
which is the same as relation (\ref{tt2}).\par 
Now, we return to the concept of temperature in gravity-induced wave function reduction. As we mentioned before in Ref \cite{RefP3} it has been argued that the nonlinear term $\frac{m \mathbf{g}^2 t^3}{\hbar} $, in the unitary transformation between the Einsteinian and Newtonian wave functions, is related to the different vacua and the Unruh effect. Now, we define a quantity corresponding to the Unruh temperature in the processes of wave function reduction, in the context of Bohmian quantum mechanics.\par 
Once again, consider a Gaussian spherical distribution in configuration space with the single degree of freedom $r$, in the short-time estimation. When the particle falls from $r = \sigma_0$ to the center of distribution( $r=0$), due to its own gravity, a volume $\frac{4}{3}\pi \sigma_0^3$ can be considered in the configuration space. The mean square velocity of the particle in the ensemble of trajectories is
\begin{equation}
\bar{\mathbf{u}^2}(\tau)=\int \mathbf{u}^2 \rho dv =\left(\mathbf{g}^2t^2\vert_{t=\tau}\right)\int_{0}^{\sigma_0} \rho dv \approx \mathbf{g}^2\tau^2
\end{equation}
Where, $\mathbf{u}=\mathbf{g}t$ is the velocity of the particle when it moves along the ith trajectory of the ensemble in its own gravity \footnote{It can be easily seen that the exact value of $\bar{\mathbf{u}^2}(\tau)$ is equal to $\displaystyle \frac{\mathbf{g}^{2} \tau^{2} ({\mathrm e}^{-\frac{1}{2}} \sqrt{2}-\mathrm{erf}(\frac{\sqrt{2}}{2}) \sqrt{\pi})}{\sqrt{\pi}}$.}. For details see Ref \cite{RefRGG3}. In the short-time estimation,  $\vert\mathbf{g}\vert=\vert\bar{\mathbf{g}}\vert$ as it has been shown in Ref \cite{RefRGG3}. Also, in this approximation the integral $\int_{0}^{\sigma_0} \rho dv=\int_{0}^{\sigma_0} \rho 4\pi r^2 dr$ has a constant finite value which we ignored because it has no effect on the final result. Now, by using the conditions $r(0)=\sigma_0$ and $r(\tau)=0$, relation (\ref{eqg}) gives the reduction time in the form
\begin{equation}\label{tt12}
\tau^2 = 2 \frac{\sigma_0}{\vert \mathbf{g}\vert}
\end{equation}
which helps us to rewrite the square mean velocity of the particle in the form
\begin{equation}\label{a2}
\bar{\mathbf{u}^2}(\tau)=2 \vert \mathbf{g}\vert\sigma_0
\end{equation}
In fact, this is the mean square velocity of the particle during the reduction time. 
Thus, relation (\ref{del}) can be written in the form
\begin{equation}
\vert \mathcal{E}(t=\tau)\vert = \frac{3}{2} m \vert\bar{\mathbf{u}^2}\vert
\end{equation}
which represents the relation between particle energy difference for two free and accelerated observers and the average velocity of the particle in the ensemble.\par 
Now, we can look at the probability distribution of the particle in the configuration space like an ideal gas composed of identical particles. Then, by using the thermodynamic relation $\bar{\mathbf{u}^2}=\frac{k_BT}{m}$ for a system with one degree of freedom(radial motion toward the center of distribution in a spherical coordinates), the kinetic energy due to the quantum effect of gravity is related to the ensemble temperature $T$ and we have
\begin{equation}\label{tem1}
\frac{k_BT}{m}=2\vert \mathbf{g} \vert \sigma_0
\end{equation}
As we know, the critical mass for the transition from the quantum world to the classical world is of the order of the Planck mass($10^{-8}Kg$), for which the reduction of the wave function  occurs \cite{RefK1,RefBassi}. On the other hand, for a particle or object with the Planck mass, the Schwarzschild radius of the particle and its Compton wavelength are the same\cite{RefLa}. If we substitute the Plank mass into  relation (\ref{Dio}), the value of $\sigma_0$ becomes about  $10^{-33}$ meters which is of the order of the Plank length. This shows that the reduction of the wave function occurs when the effects of quantum gravity are important.
According to these arguments, we take the matter distribution radius (the characteristic width of the wave packet) equal to the Compton wavelength of the particle, i.e. $\sigma_0 \sim \lambda_C = \frac{\hbar}{m}$. Then, relation (\ref{tem1}), gives:
\begin{equation}\label{u1}
T = \frac{\hbar \vert \mathbf{g} \vert}{k_B}
\end{equation}
which  corresponds to the Unruh temperature in quantum field theory in the curved spacetime\footnote{If we write the Compton wavelength in the form $\lambda_C = \frac{\hbar}{mc}$, we will have $T = \frac{\hbar \vert \mathbf{g} \vert}{k_B c}$. In non-relativistic regime, we can consider $c \longrightarrow \infty$. Then, this temperature does not have a significant value in non-relativistic regime.}. It may be appropriate to call it "reduction temperature". This relation is derived for a particle with mass of the order of the Planck mass and the width of its wave packet close to the Schwarzschild radius. Then, if we substitute relations $\sigma_0 = 2Gm$ and $\vert \mathbf{g}\vert=\frac{Gm}{\sigma_0^2}$ into relation (\ref{u1}), we get the reduction temperature in the form 
\begin{equation}\label{ha}
T = \frac{\hbar}{k_B Gm}
\end{equation}
which is of  order of the Hawking temperature\footnote{Because of the Gaussian form of the matter distribution, relations (\ref{u1}) and (\ref{ha}) are obtained by approximation. For example, the value of the gravitational acceleration for the Gaussian distribution is $\vert \mathbf{g} \vert=2\sqrt{\frac{2}{\pi}}\frac{Gm}{\sigma_0^2}$, whereas we simply set it equal to $\frac{Gm}{\sigma_0^2}$. See Ref \cite{RefRGG3} }. In other words, it is the Unruh temperature when the radius of distribution is $\sigma_0=2Gm$ in the configuration space. We expected these results. Because, both in the quantum field theory in a curved spacetime and in the gravity-induced wave function reduction, acceleration or gravity transforms a pure quantum state into a mixture of states. This mixture can be seen as a thermodynamic system with a specific temperature. \par 
Relations (\ref{u1}) and (\ref{ha}) were obtained, through the concept of geometrization or gravitization of quantum mechanics in the Bohmian context. But, what is very important and  mysterious is the hidden physical concept behind these phenomena. A pure quantum state for an observer, seems to another observer as a mixture of states with a specific temperature. Just, like a thermodynamic system with a specific temperature. If we give originality to geometry, equivalence principle and determinism, then we can conclude that the statistical behaviors in quantum mechanics is not intrinsic. Rather, like a  thermodynamic system, it refers to our ignorance of all the necessary information about the system. In other words, there may be an underlying fundamental deterministic level whose information manifest to us as the probabilistic rules of quantum mechanics. It may be related to the issue of hidden variables. But, the hidden variables are not known to us to fully describe the quantum systems. It seems that quantum problems and gravity meet where the concept of hidden variables becomes prominent. This is an open problem and needs further study and research.
\section{Conclusion}
The results of this research can be divided into two parts.\\
In the first part, we used the equivalence principle of general relativity to describe the gravity-induced wave function reduction in the Bohmian quantum mechanics. In this regard, we studied the quantum motion of the particle in its own gravity in a short time estimation. In a short-time estimation ($\sigma \approx \sigma_0$), the average of gravitational and quantum accelerations can be considered constant\cite{RefRGG3}. We used this point that the nature of the quantum force is like the nature of an inertial force \cite{RefMach}. In other words, the dynamics of the particle under the influence of the quantum force is equivalent to the dynamics of a particle in a non-inertial frame with the quantum acceleration. So, to have a free motion in this accelerated frame and to establish the third statement of WEP, the acceleration of the frame should be equal to the self-gravitational acceleration. This led us to obtain the famous criterion of Diosi for the wave function reduction. Thus, by using the concept of trajectories and dynamics of the particle we concluded that the establishment of the third statement, leads to the wave function reduction systematically. \par 
In the second part, we obtained the reduction time of the wave function by using the concepts of Einsteinian and Newtonian observers, which has been introduced in Ref \cite{RefP3}. Here, we used the energy formula $E=-\frac{\partial S}{\partial t}$ which relates the energy of the particle to the phase of the wave function. This relation helped us to obtain the uncertainty in particle energy to estimate the reduction time. This can be seen as a new approach in this context. But an interesting thing that happened in this part, was the obtaining of a temperature corresponding to the Unruh temperature. In Ref \cite{RefP3}, it has been argued that in the problem of wave function reduction there is a temperature corresponding to the Unruh temperature of quantum field theory in a curved spacetime. We obtained a  relation for this temperature through the capabilities of the Bohmian quantum mechanics. This has been done for the first time and can be interesting in its own way. \par 
Finally, we argued that the connection between  gravity and quantum mechanics, in our discussion, is not accidental. Rather it can refer to deeper concepts.






\bibliography{sn-bibliography}

\end{document}